# A Multi-chain Measurements Averaging TDC Implemented in a 40 nm FPGA

Qi Shen, Shubin Liu, Binxiang Qi, Qi An, Shengkai Liao, Chengzhi Peng, Weiyue Liu

*Abstract*—**A high precision and high resolution time-to-digital converter (TDC) implemented in a 40 nm fabrication process Virtex-6 FPGA is presented in this paper. The multi-chain measurements averaging architecture is used to overcome the resolution limitation determined by intrinsic cell delay of the plain single tapped-delay chain. The resolution and precision are both improved with this architecture. In such a TDC, the input signal is connected to multiple tapped-delay chains simultaneously (the chain number is M), and there is a fixed delay cell between every two adjacent chains. Each tapped-delay chain is just a plain TDC and should generate a TDC time for a hit input signal, so totally M TDC time values should be got for a hit signal. After averaging, the final TDC time is obtained. A TDC with 3 ps resolution (i.e. bin size) and 6.5 ps precision (i.e. RMS) has been implemented using 8 parallel tapped-delay chains. Meanwhile the plain TDC with single tapped-delay chain yields 24 ps resolution and 18 ps precision.**

## I. INTRODUCTION

Over the past, the performance of FPGA based TDC has been improved rapidly from 200 ps resolution [1] down to below 10 ps resolution [2], which mainly benefits from both the improved FPGA manufacturing process technology and optimized or novel proposed TDC architecture. In 1997, Kalisz et al. implemented a 200 ps resolution TDC on a 0.65 um CMOS FPGA from QuickLogic [1]. They used a vernier delay line method to achieve time interpolation. In 2003, a TDC with 400 ps bin size was implemented by Wu et al in an Altera ACEK FPGA. The time interpolation technique with tapped-delay line (TDL) was used with cascade chains [4]. In 2006, Song et al. proposed a novel TDC architecture using dedicated carry lines of an FPGA as TDL to perform time interpolation [5]. In 2008, a novel wave union architecture was proposed by Wu et al, which make multiple measurements with a single TDL [2]. The resolution and precision can both be further improved beyond the intrinsic cell delay of the TDL by the wave union method. The Cyclone II FPGA was used with 16 time measurements to get 10 ps RMS precision. Wang et al. make a theoretical analysis of the wave union TDC and present an improved wave union scheme implemented in Xilinx Virtex 4 FPGA [6]. In 2010, M. Daigneault et al. proposed a multiple parallel TDLs based TDC architecture, the 24 ps RMS and 10 ps bin size was obtained using 10-parallel TDLs in Virtex II Pro FPGA [7]. In 2011, they also implemented a 10 ps precision TDC using the dynamic reconfiguration function and calibration process in the same FPGA[8].

The TDL based interpolation architecture has proven to be most appropriate to implement an FPGA TDC. The multiple time measurements averaging method can be used to improve the TDC performance, such as the wave union method. However, the dead time of the wave union method could be increased. This paper presented an FPGA TDC implemented in a 40 nm Virtex 6 FPGA using multi-chain measurements averaging architecture. Because the parallel multi-chains measure the input signal simultaneously, the dead time is not increased. The remaining part of the paper is organized as follows. The methodology and architecture of the proposed TDC is explained in Section II. In Section III, we show bench-top test results. A conclusion is given in Section IV.

## II. METHODOLOGY

The block diagram of the multi-chain measurements averaging TDC is shown as Fig. 1. A TDC channel is consisted of M (M>1) tapped-delay chains. Each tapped-delay chain is just a plain TDC, using the dedicated carry-lines of FPGA. There is a delay cell between every two adjacent chains. The delay cell is also a dedicated carry-chain unit in our design. Each chain should generate a TDC time for a hit input signal, so totally M TDC time values should be got for a single hit input.

Manuscript received May 22, 2014. This work was supported in part by the Knowledge Innovation Program of the Chinese Academy of Sciences (KJCX2-YW-N27) and in part by the National Natural Science Foundation of China.

Qi Shen is with the State Key Laboratory of Particle Detection and Electronics, USTC; Hefei National Laboratory for Physical Sciences at Microscale and Department of Modern Physics, University of Science and Technology of China, Hefei, Anhui 230026, China; and Shanghai Branch, CAS Center for Excellence and Synergetic Innovation Center in Quantum Information and Quantum Physics, University of Science and Technology of China, Shanghai, 201315, China (e-mail: shenqi@ustc.edu.cn).

Binxiang Qi, Shubin Liu, and Qi An are with the State Key Laboratory of Particle Detection and Electronics, USTC; CAS Center for Excellence and Synergetic Innovation Center in Quantum Information and Quantum Physics; and Department of Modern Physics, University of Science and Technology of China, Hefei, Anhui 230026, China (e-mail: liushb@ustc.edu.cn).

Shengkai Liao, Chengzhi Peng are with the Hefei National Laboratory for Physical Sciences at Microscale and Department of Modern Physics, University of Science and Technology of China, Hefei, Anhui 230026, China; and Shanghai Branch, CAS Center for Excellence and Synergetic Innovation Center in Quantum Information and Quantum Physics, University of Science and Technology of China, Shanghai, 201315, China (e-mail: skliao@ustc.edu.cn).

Weiyue Liu is with the College of Information Science and Engineering, Ningbo University, Ningbo, China.

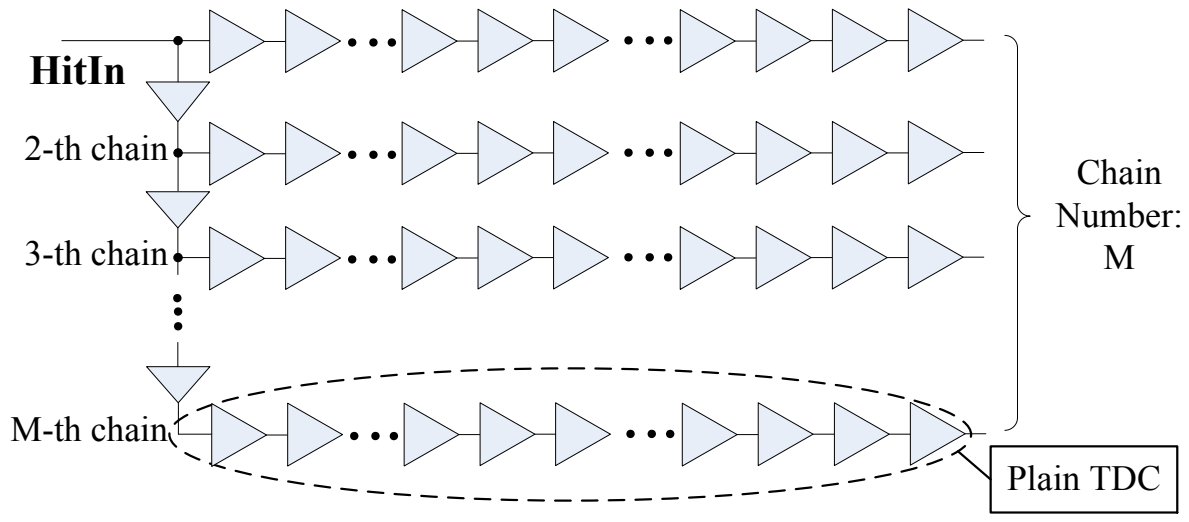

Fig. 1. Block diagram of the multi-chain measurements averaging TDC.

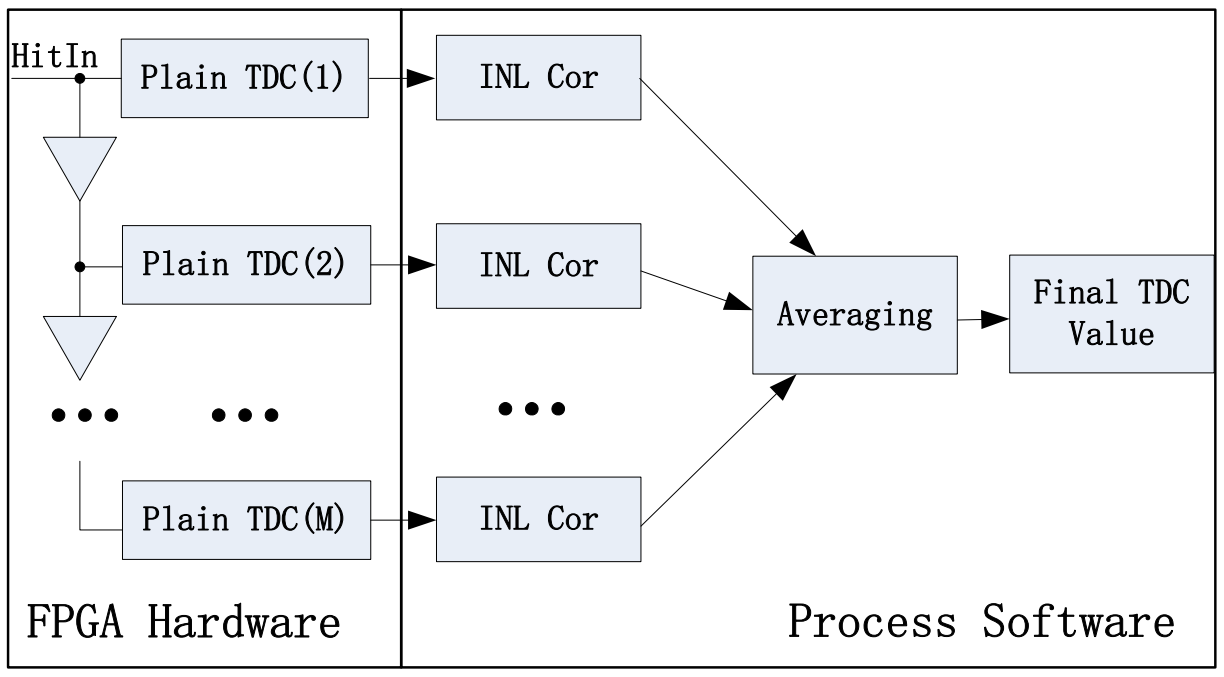

Fig. 2. Implementation of multi-chain measurement averaging TDC.

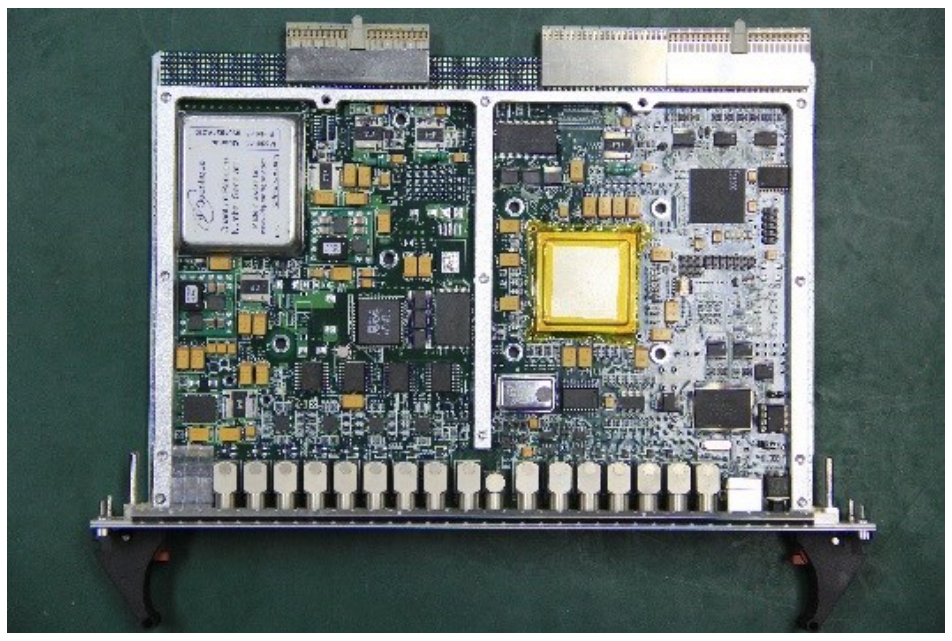

Fig. 3. A picture of the evaluation board based on 6U CPCI standard

The implementation of the whole multi-chain averaging TDC is depicted in Fig. 2, which include FPGA hardware and data process software. The data process software is implemented by the Matlab tool in our design. Note that the software process can also be done in FPGA easily. Each integral non-linearity (INL) for all M tapped-delay chains should be calculated using a code-density test. Each raw TDC code can be corresponded to a plain TDC time value according to these INLs, this process is often called INL calibration. For a hit signal, there should be M time values $t_{Plain}(\mathrm{i})$ obtained, where $1 \leq \mathrm{i} \leq \mathrm{M}$.

The averaging delay time $T_D(\mathrm{m})$ between each other chain and the first chain should be calculated using the plain TDC time data, where the $\mathrm{T}_D(\mathrm{m})$ ( $2 \leq \mathrm{m} \leq \mathrm{M}$ ) means the averaging delay time between the m-th chain and the first chain. $\mathrm{T}_D(1)$ is set to zero for the convenience of calculation.

The final TDC time for the hit signal is $\mathrm{t}_{Final}$, which should be calculated by averaging all M time values as

$$\mathrm{t}_{Final} = \frac{1}{M}\sum_{i=1}^{M}(t_{Plain}(\mathrm{i}) - \mathrm{T}_D(\mathrm{i})).$$

Fig. 3 shows the photograph of our evaluation board of the multi-chain measurements averaging TDC implemented in XC6VLX240T-2FFG1156 device from Xilinx Virtex-6 family. The circuit board is a standard 6U CPCI module.

## III. Test Results

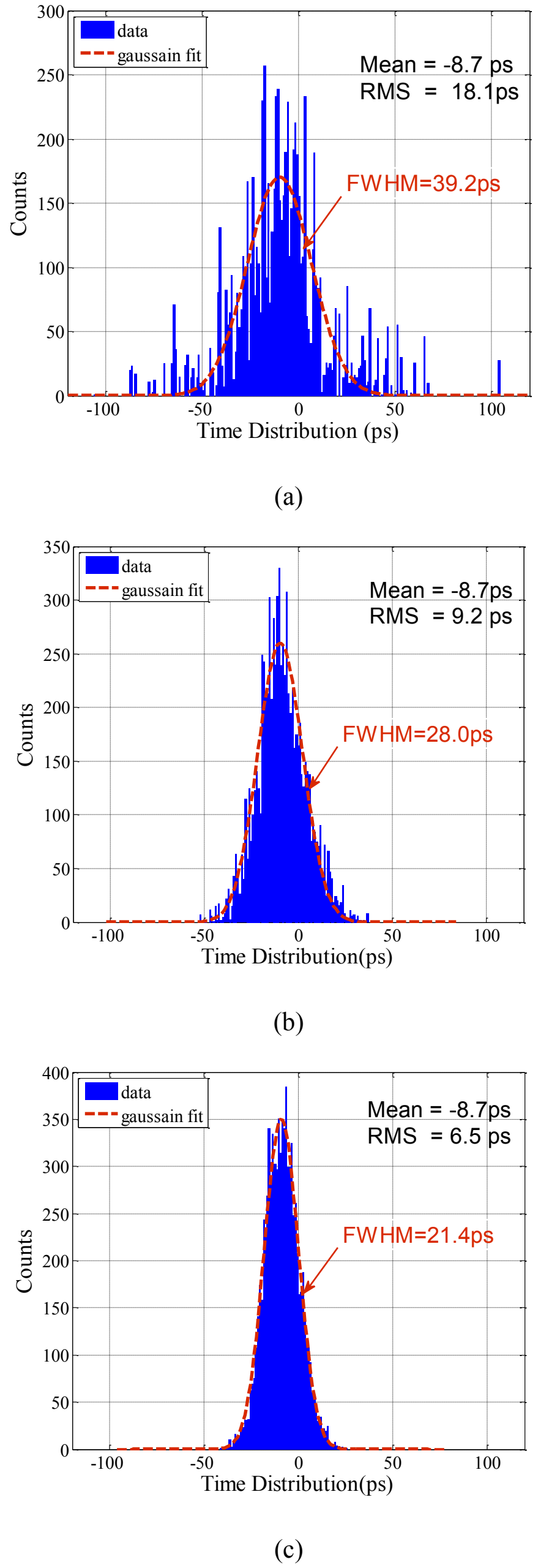

Fig. 4. Typical timing precision RMS test result of the TDC with different chain number M. (a) Chain number M = 1 (plain TDC); (b) Chain number M=4; (c) Chain number M=8.

A 2-channel multi-chain measurements averaging TDC is implemented with 8 parallel chains (i.e. chain number M = 8) for each channel. Each chain forms a common plain TDC based on the dedicated carry lines, which is named CARRY4 block [10] in Vertex 6 FPGA, which is the same architecture as Vertex 5 family. In order to get better linearity performance, the CARRY4 is subdivided as 2 delay taps [10]. Note that the CARRY4 can actually be subdivided into 4 delay taps, but with worse non-linearity. Besides the multi parallel chains for fine time measurement, clock counters are used for coarse time measurement. The clock frequency is 160 MHz. The 6.25 ns period corresponds to approximately 260 delay taps for each chain. The actual length of each tapped delay line is set to be 276 in order to allow for technological spread and temperature drift.

The cable delays measurements method was used to evaluate the overall precision [5], which is the most important parameter of a TDC. Fig. 4 shows the typical statistical spread of the time precision of the time measurement with different chain number value M. In Fig. 4(a) the timing precision of a single TDC channel is about 18 ps RMS with M=1, which means a plain TDC. From Fig. 4(b) and Fig. 4(c) we can see that the RMS is improved to 9.2 ps and 6.5 ps with M=4 and M=8, respectively. Note that the RMS values above were the results of the measured values divided by $\sqrt{2}$, as the measured values of standard deviation contain two TDC channels. The relationship between RMS timing precision and chain number M is shown in Fig. 5, in which the RMS value is decreased with M increased.

As seen in Fig. 6, the bin width is effectively subdivided as the chain number M increased. The mean bin size of plain TDC (M=1) is 24 ps, while the mean bin size is reduced to 2.93 ps when M=8.

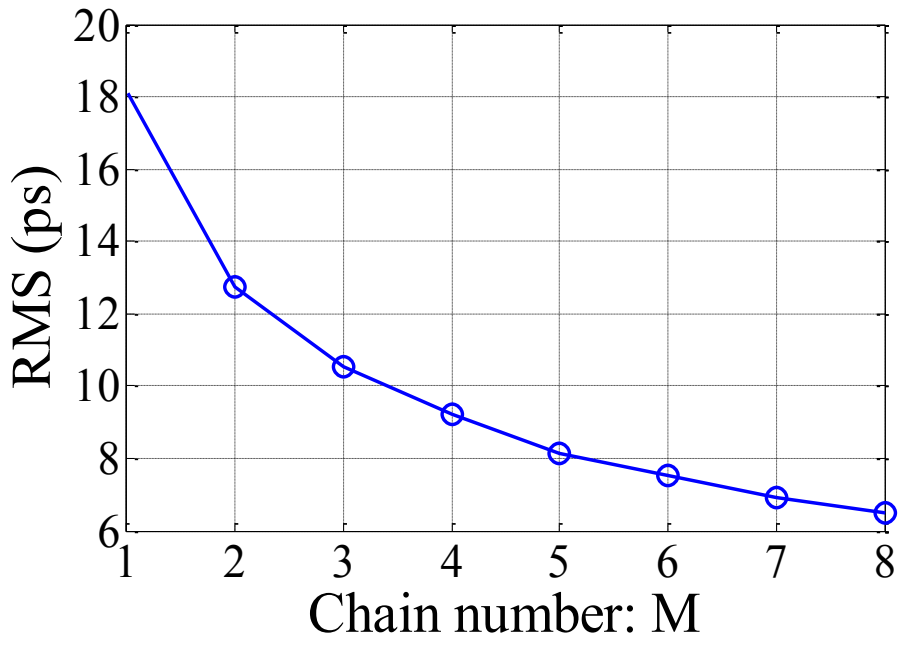

Fig. 5. Relationship between RMS timing precision and chain number M.

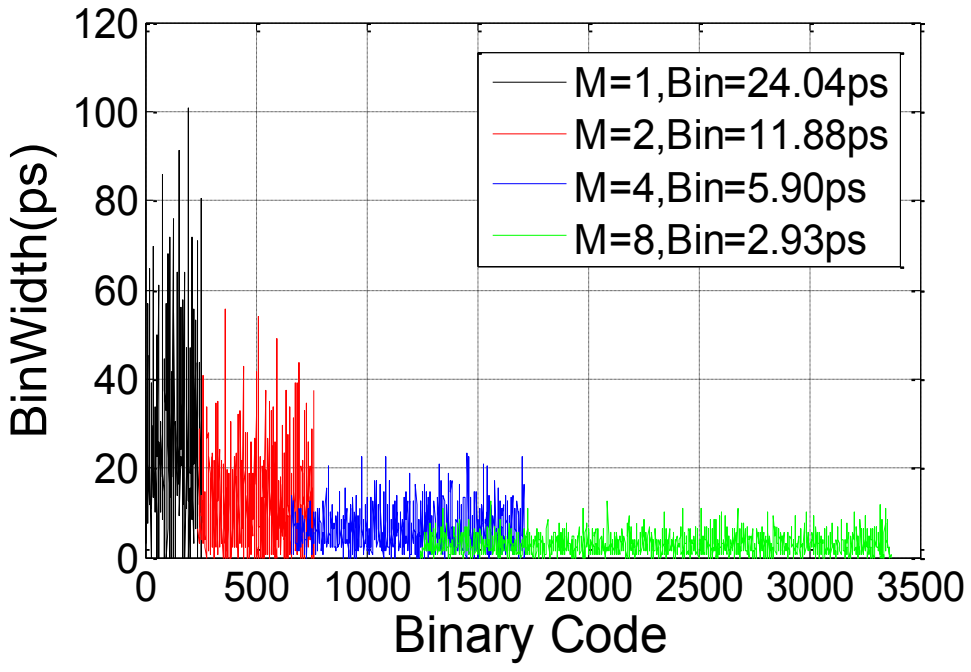

Fig. 6. Bin width histograms with different chain number M.

The statistical code density test was used to perform the characterization of the non-linearity of the proposed TDC. Fig. 7(a) and Fig. 7(b) show the differential non-linearity (DNL) and integral non-linearity (INL) with M = 8, respectively. The red lines represent the results with the original binary fine codes, with no calibration, while the results after INL calibration are shown in the blue lines. The calibration of each tapped delay chain translates the binary fine codes to times, which is carried out before the calculating of offset delay between delay chains and averaging process. The results show that the DNL is not improved by calibration. However, the INL is effectively improved by calibration, which is reduced to the range of (-9.8, 6.2) LSB from the range of (-12.5, 10.7) LSB; here the LSB is 2.93 ps.

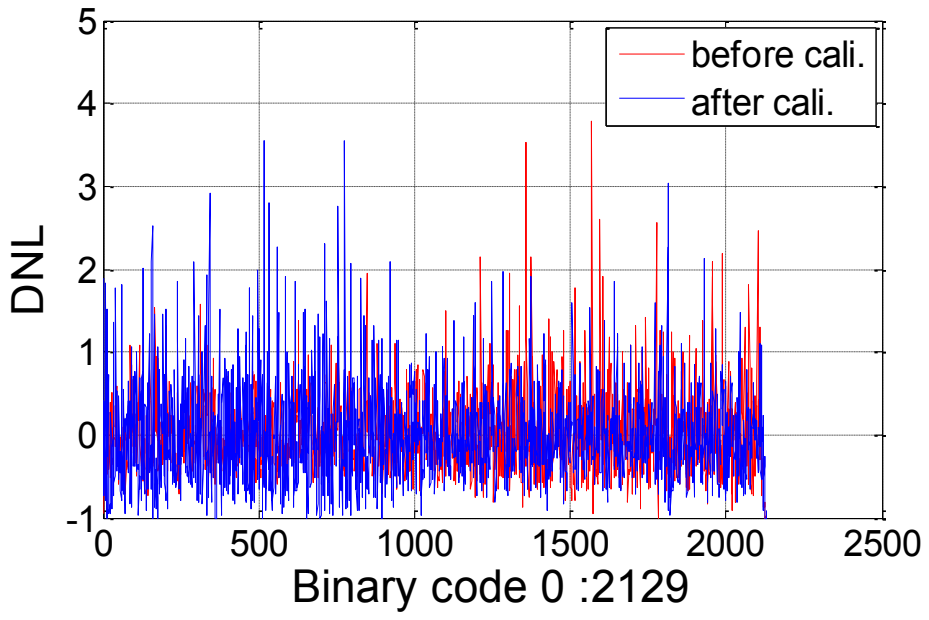

(a)

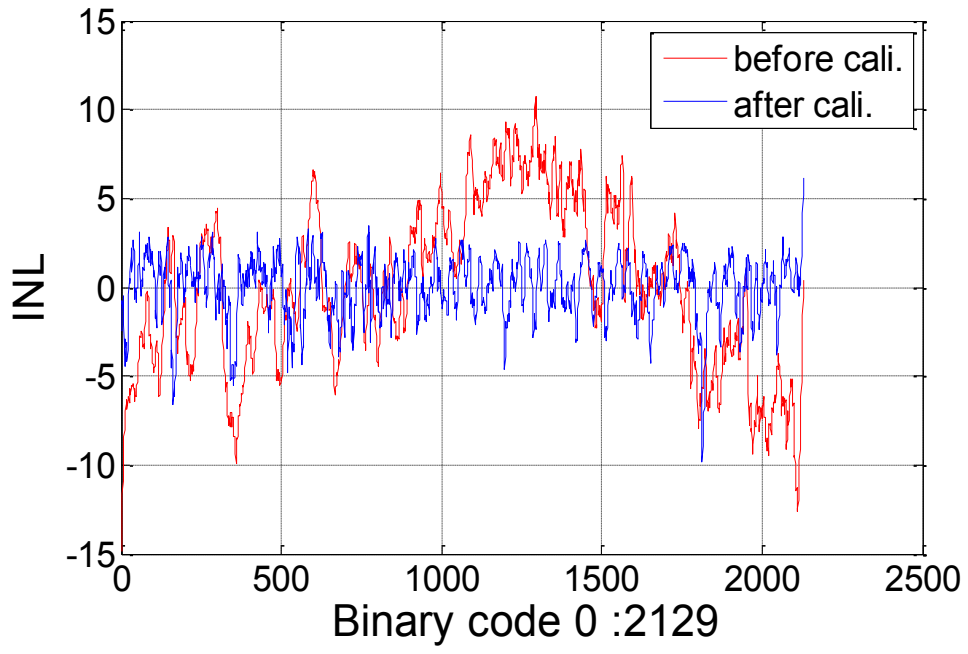

(b)

Fig. 7. Nonlinearity performance of the multi-chain averaging TDC with M = 8, where (a) is Differential Nonlinearity, (b) is Integral Nonlinearity. The red lines represent the results before calibration and the blue lines represent the results after calibration of each tapped delay chains.

In fact, the TDC output code pattern is actual time value which could be floating number rather than binary integral number caused by the calibration and averaging process of the proposed TDC. We can set the bin size or LSB manually [7] with different values which will results in different non-linearity. The non-linearity will be better with bigger LSB setted. The DNL and INL of the multi-chain averaging TDC (M = 8) and the plain TDC (M = 1) with the same LSB (24 ps) is shown in Fig. 8 for comparison. Compare to plain TDC, the multi-chain averaging TDC can effectivly improve the non-linearity. The DNL and INL are reduced from (-1, 2.8) LSB and (-3.7, 3.3) LSB to (-0.7, 0.8) LSB and (-1, 0.7) LSB, respectively.

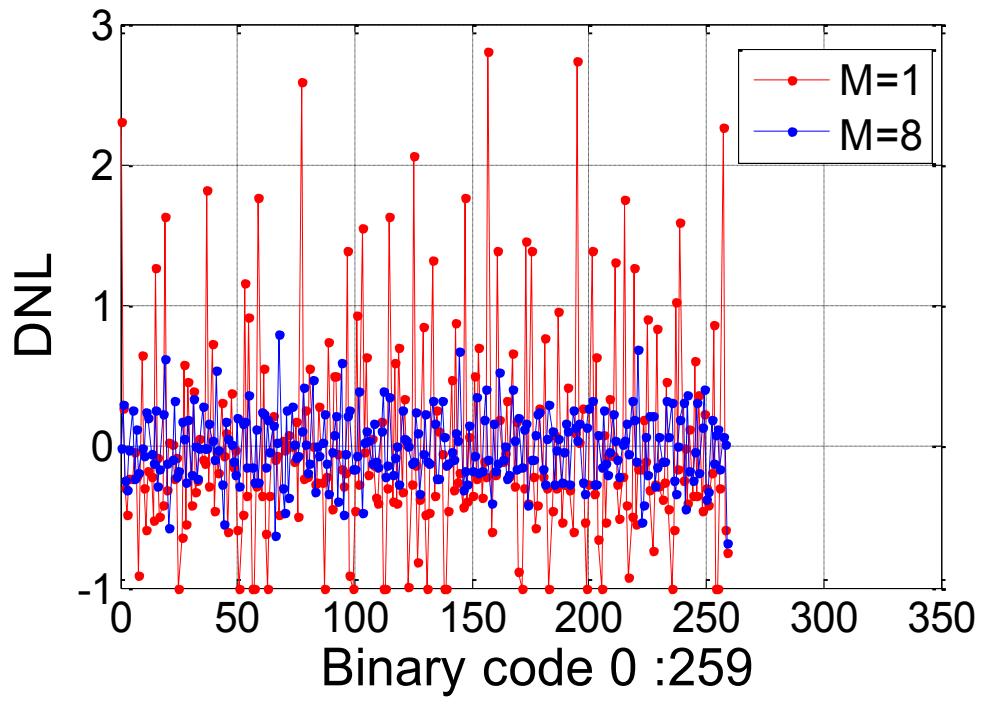

(a)

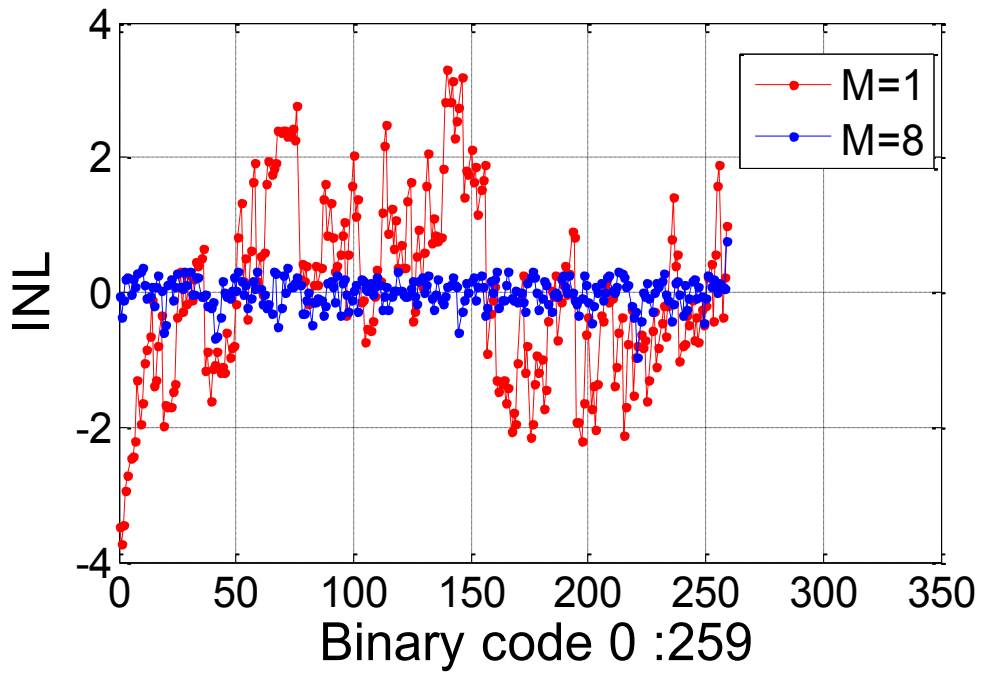

(b)

Fig. 8. Non-linearity of the multi-chain averaging TDC (M = 8) and the plain TDC (M = 1) with the same LSB 24 ps. (a) Differential Non-linearity; (b) Integral Non-linearity.

## IV. Conclusion

A 2-channel multi-chain measurements averaging TDC is implemented with maximum 8 parallel tapped delay chains in Vertex 6 FPGA. Test results indicate that the precision and resolution can both be improved. The best precision with M = 8 is 6.5 ps RMS while the plain TDC is 18 ps RMS.

## References

[1] J. Kalisz, R. Szplet, J. Pasierbinski, and a. Poniecki, “Field-programmable-gate-array-based time-to-digital converter with 200-ps resolution,” *IEEE Trans. Instrum. Meas.*, vol. 46, no. 1, pp. 51–55, 1997.

[2] J. Wu and Z. Shi, “The 10-ps wave union TDC: Improving FPGA TDC resolution beyond its cell delay,” in *IEEE Nuclear Science Symposium Conference Record*, 2008, pp. 3440–3446.

[3] M. Fries and J. Williams, “High-precision TDC in an FPGA using a 192 MHz quadrature clock,” in *IEEE Nuclear Science Symposium Conference Record*, 2002, vol. 1, pp. 580–584.

[4] J. Wu, Z. Shi, and I. Wang, “Firmware-only implementation of time-to-digital converter (TDC) in field-programmable gate array (FPGA),” *IEEE Nucl. Sci. Symp. Conf. Rec.*, vol. 1, pp. 177–181, 2003.

[5] J. Song, Q. An, and S. Liu, “A high-resolution time-to-digital converter implemented in field-programmable-gate-arrays,” *IEEE Trans. Nucl. Sci.*, vol. 53, no. 1, pp. 236–241, Feb. 2006.

[6] J. Wang, S. Liu, L. Zhao, X. Hu, and Q. An, “The 10-ps Multitime Measurements Averaging TDC Implemented in an FPGA,” *IEEE Trans. Nucl. Sci.*, vol. 58, no. 4, pp. 2011–2018, Aug. 2011.

[7] M.-A. Daigneault and J. P. David, “A novel 10 ps resolution TDC architecture implemented in a 130nm process FPGA,” *Proc. 8th IEEE Int. NEWCAS Conf.*, pp. 281–284, Jun. 2010.

[8] M. Daigneault and J. David, “A High-Resolution Time-to-Digital Converter on FPGA Using Dynamic Reconfiguration,” *IEEE Trans. Instrum. Meas.*, vol. 60, no. 6, pp. 2070–2079, 2011.

[9] M. W. Fishburn, L. H. Menninga, C. Favi, and E. Charbon, “A 19.6 ps, FPGA-Based TDC With Multiple Channels for Open Source Applications,” *IEEE Trans. Nucl. Sci.*, pp. 1–1, 2013.

[10] L. Zhao, X. Hu, S. Liu, J. Wang, Q. Shen, H. Fan, and Q. An, “The Design of a 16-Channel 15 ps TDC Implemented in a 65 nm FPGA,” *IEEE Trans. Nucl. Sci.*, vol. 60, no. 5, pp. 3532–3536, Oct. 2013.